\documentclass[aps,prl,showpacs,twocolumn,floatfix]{revtex4}
\usepackage{times}
\usepackage{graphicx,color}
\usepackage{amsmath,amssymb}
\def\beq{\begin{equation}}
\def\eeq{\end{equation}}
\def\bea{\begin{eqnarray}}
\def\eea{\end{eqnarray}}

\begin{document}

\title{Predicting the coherence resonance curve using a semi-analytical 
treatment}

\author{Santidan Biswas} 
\email{santidan@phy.iitb.ac.in}

\author{Dibyendu Das}
\author{P. Parmananda}
\author{Anirban Sain}

\affiliation{Department of Physics, Indian Institute of Technology Bombay,
  Powai, Mumbai-400 076, India.}
%\author{Dibyendu Das}
%\affiliation{Department of Physics, Indian Institute of Technology, Bombay,
%  Powai, Mumbai-400 076, India.}
%\author{Anirban Sain}
%\affiliation{Department of Physics, Indian Institute of Technology, Bombay,
%  Powai, Mumbai-400 076, India.}
%\author{P Parmananda}
%\affiliation{Department of Physics, Indian Institute of Technology, Bombay,
%  Powai, Mumbai-400 076, India.}
\date{\today}

\begin{abstract}
Emergence of noise induced regularity or Coherence Resonance in
nonlinear excitable systems is well known. We explain theoretically
why the normalized variance ($V_{N}$) of inter spike time intervals, which
is a measure of regularity in such systems, has a unimodal profile.
Our semi-analytic treatment of the associated spiking process produces
a general yet simple formula for $V_{N}$, which we show is in very good
agreement with numerics in two test cases, namely the FitzHugh-Nagumo
model and the Chemical Oscillator model.
\end{abstract}

%{polymer solutions,probability - stochastic process}

\maketitle
\section{Introduction}
Many deterministic, nonlinear, excitable systems, for example, the 
FitzHugh-Nagumo model (FHN) \cite{Pikovsky} or the Chemical Oscillator model 
(CO) \cite{Karantonis}, undergo bifurcation from a stable focus to a stable  
limit cycle (LC) behavior when a system parameter is tuned. However,
holding the parameter near the bifurcation point, on the stable focus side
the system can still be made to exhibit spiking behavior (which is 
otherwise the signature of a limit cycle), by adding a random 
uncorrelated noise to the system. The noise forces the system to 
intermittently jump across the bifurcation point in the parameter 
space. As a result of these random excursions, the system exhibits 
intermittent cyclic behavior which manifests as spikes in the dynamical 
variable. Interestingly, the time intervals $\tau_{p}$, between two 
successive noise driven spikes, which are in general irregular, strangely 
becomes fairly regular at an optimal noise value (the resonance point). 
This phenomenon is called Coherence Resonance. It has attracted considerable 
interest theoretically as well as experimentally \cite{Benzi,Benzi2,Nicolis,
Nicolis2,Gammaitoni,Neiman,Lindner,Parmananda}, as quite 
counter-intuitively order arises with the aid of tuned randomness.  
A quantitative means of detecting this resonance point is enumerating the 
normalized variance ($V_{N}$) defined by
%\begin{equation}
%\mathcal{R} 
%$V_{N}$ = \frac{\sqrt{{\langle \tau_{p}^2 \rangle} - 
%{\langle \tau_{p} \rangle}^2}}{\langle \tau_{p} \rangle}, 
%\label{snr}
%\end{equation}
$V_{N} = {\sqrt{{\langle \tau_{p}^2 \rangle} - 
{\langle \tau_{p} \rangle}^2}}/{\langle \tau_{p} \rangle}$, 
as a function of noise strength. Here $\langle . \rangle$ denotes 
statistical time average. Typically $V_{N}$ is enumerated from time-series analysis
of spikes generated by the system, subjected to noise.  The noise
strength at which minimum of $V_{N}$ occurs is the desired point of
resonance.

The analytical work so far on this subject, have either dealt with a
toy model \cite{Pikovsky}, or addressed special limits of the FHN
model e.g, very weak noise \cite{Lacasta}, and infinite time scale
separation between the fast and slow variables
\cite{Lindner,Lindner1,Lindner2}.  A pioneering qualitative understanding of
the phenomenon is given by Pikovsky and Kurths \cite{Pikovsky}, who
argue that the resonance happens as a competition between two time
scales -- the activation time $t_{a}$ (the time between the end of one
spike and beginning of another) and the excursion time $t_{e}$ i.e.,
duration of a spike. The inter spike interval (ISI) $\tau_{p} = t_a +
t_e$. They claim that $t_a$ has a strong dependence on noise intensity
and follows a simple Kramer's \cite{kramers} like formula, whereas
$t_e$ has a much weaker noise dependence and corresponds to the decay
time of unstable excited state.  Kramers theory describes the noise
driven escape time $\tau_{\rm esc}$ of a particle (say $y$) from a
deep potential trap, and gives $\tau_{\rm esc}\sim \exp(E_b/D^2)$;
here $D$ is noise amplitude, and $E_b$ is the barrier height. But
excitable systems with two coupled variables $x$ and $y$ pose new
challenges: the barrier $E_b$ is both dynamic and $D$ dependent.  The
effective barrier for $y$ is dynamic as it is generated by $x$ which
itself is a dynamical variable.  Furthermore our numerical studies show
that barrier parameters, like its width $\delta$, are indeed $D$
dependent. In this paper, we avoid invoking Kramers picture apriori,
and show that the timescales $t_e$ and $t_a$ can be understood from
alternative arguments.

%Also in
%\cite{Pikovsky} moments of $\tau_p$ were obtained from a ``first
%passage process'' rather than a Kramer's escape (on which their
%qualitative argument was based).
%Interestingly, for a completely opposite phenomenon (in small $D$ limit) called
%anti-coherence resonance,  \cite{Lacasta} gave a theoretical formula
%for $V_{N}$ in the FHN model, but again all parameters of their formula had
%to be finally fixed by time series analysis.

We derive below a simple theoretical formula for $V_{N}$, which 
will be generally applicable to any nonlinear system exhibiting 
coherence resonance. There are parameters in the universal formula, which 
depend on the specific details of the nonlinear system at hand, and can
only be fixed by some amount of numerical or alternatively experimental
analysis. Thus the formula is semi-theoretical. Although this may seem 
as no less work than the usual time-series analysis, as we show below, 
it certainly involves incorporation of enhanced understanding of the 
phenomenon compared to what existed before. To support our claim of
generality, we study  two very different nonlinear systems: the 
FHN model \cite{Pikovsky} and the 
CO model \cite{Karantonis,Parmananda,Santos}. We show that our 
predicted formula fits quite well, with the curve of $V_{N}$ obtained by brute 
force time-series analysis, in both the cases.

\section{Model}
Before starting our main analysis, let us define the FHN and CO systems 
in the presence of noise, to make this paper self contained. 
The FHN model has the following equations
\begin{eqnarray}
\epsilon \frac{dx}{dt} = x - \frac{x^3}{3} - y, ~~~   
\frac{dy}{dt} = x + a + D\xi(t).
\label{fhn2}
\end{eqnarray}
Here %$x$ and $y$ are the dynamical variables, 
$a$, $D$ and 
$\epsilon$ ($\ll 1$) are the three parameters.  For $|a| > 1$, there 
is a stable fixed point at $x_{*} = -a$ , $y_{*} =
\frac{a^3}{3} -a$, while for $|a| < 1$ a limit cycle exists in the
$x-y$ space and dynamics of both the variables are periodic. The value 
of $a$ on the fixed point side, which we hold fixed for our simulation,
is denoted by $a_{0}$. The parameter $D$ is the amplitude of the Gaussian 
white noise $\xi$, for which $\langle \xi(t) \rangle=0$ and 
$\langle \xi(t) \xi(t^{'})\rangle =  \delta(t-t^{'})$.
The small parameter $\epsilon$ makes the motion on the limit 
cycle much faster along the $x$ direction than the $y$. 
The second model of CO is defined by the following equations:
\begin{eqnarray}
\epsilon \frac{du}{dt} &=& \frac{v-u}{R} - f(u,c), \nonumber \\
%\label{eqn1} \\
\frac{dc}{dt} &=& \frac{u-v}{R} +(1-c) + \alpha f(u,c)
\label{eqn2}
\end{eqnarray}
where 
%\begin{equation}
$f(u,c) = c(a_{1}u + a_{2}u^{2} + a_{3}u^{3})$
%\label{eqn3}
%\end{equation}
and
%\begin{equation}
$v = v_{0} + D \xi(t)$. 
%\label{eqn4}
%\end{equation} 
Here $u$ and $c$ are the dynamical variables and $R$, $v_0$, $a_1$, $a_2$, 
$a_3$, $\epsilon$, $\alpha$, and $D$ are the parameters. $v$ is the bifurcation parameter. Limit cycle exists for the values $ v \le 29.235$ whereas
for $v > 29.235$, a steady state fixed point behavior is observed. The system 
variables and parameters are derived from the reaction-rate kinetics of the 
interacting chemical species. The details regarding the construction of the 
model equation are furnished elsewhere \cite{Karantonis,Parmananda,Santos}.

\begin{figure}
\includegraphics[scale=0.7]{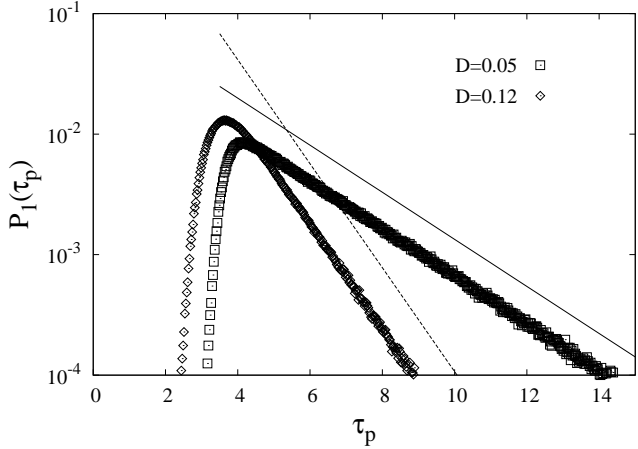}
\caption{Semi-log plot of $P_{1} (\tau_{p})$ against $\tau_p$ 
for FHN model. Here $a_{0} = 1.1$ and $\epsilon = 0.01$ (see Eq. 
(\ref{fhn2})) --- these same values are used for other
figures in the paper. The straight lines are exponential 
fits (placed higher for visual clarity) to the tail to obtain 
$t_{\rm esc}^{\rm num}$ values and they are $2.225$, $1.007$ for 
$D = 0.05$ and $0.12$ respectively. The corresponding 
$t_{\rm min}^{\rm num}$ (defined in the text) values are $3.573$ and 
$2.904$. }
\label{probdist}
\end{figure}

\section{Results}
If one makes a simple-minded first guess that the inter spike intervals  
$\tau_p$ have a Poisson distribution, then $V_{N}$ would be a constant 
(independent of noise strength) which is empirically not the case. 
So what is the distribution of $\tau_p$? For a random train of spikes
which are almost independent, it seems very likely that the distribution 
of ISI will have an exponential tail \cite{gardiner}. 
Yet a specialty of the spikes in 
the non-linear systems of our concern, is that a new spike cannot 
arise until the last spike subsides. Thus $\tau_p$ cannot be any smaller 
than characteristic `spike width' $w_s$ (a finite quantity), 
i.e. the distribution of $\tau_p$ is expected to have a sharp lower cutoff 
at some finite $\tau_{\rm min}$. We stress here that if this lower cutoff 
were absent, then $V_{N}$ would have had no variation and coherence resonance
would have vanished. Thus we expect the 
probability density of $\tau_p$ to be,       
\begin{equation}
P_{1} (\tau_{p}) = N \Theta(\tau_{p} -\tau_{\rm min}(D)) 
\exp\left(-\frac{\tau_{p}} {\tau_{\rm esc}(D)}\right).
\label{distri_taup} 
\end{equation}
Here $\Theta$ is the Heaviside function \cite{arfken}, while 
$\tau_{\rm esc}$ is the
characteristic time associated with exponential tail of $P_{1}
(\tau_{p})$. In Eq.~(\ref{distri_taup}), the normalization constant 
$N = {(e^{\tau_{\rm min} / \tau_{\rm esc}})}/{\tau_{\rm esc}}$. 
We have checked that the distribution of $\tau_p$
obtained from the time series analysis of the FHN and CO models are
consistent with Eq.~(\ref{distri_taup}) --- see Fig.~\ref{probdist} for
numerically obtained $P_{1} (\tau_{p})$ for the FHN system for
two different $D$ values. Despite the two values of $D$, one being 
away and another close to the resonance point, one can see
clearly that the shape of the curves $P_{1} (\tau_{p})$ shows no
qualitative variation. Of course the quantities $\tau^{\rm num}_{\rm
min}$ and $\tau^{\rm num}_{\rm esc}$ (where the superscript ``num''
denotes numerical) are functions of $D$; in fact both decrease with
$D$. The notational distinction between $\tau^{\rm num}_{\rm min}$ in
Fig.~\ref{probdist} and $\tau_{\rm min}$ in Eq.~(\ref{distri_taup}) is
necessary, as the numerical curve in Fig.~\ref{probdist} does not rise
strictly as a $\Theta$ function. To be precise, in
Fig.~\ref{probdist}, $\tau^{\rm num}_{\rm min}$ is defined as the
average of the time $\tau^{\rm num}_{\rm {min,l}}$ at which $P_{1}
(\tau_{p})$ just starts becoming nonzero and the time $\tau^{\rm
num}_{\rm {min,r}}$ at which $P_{1} (\tau_{p})$ reaches a peak. On
the other hand $\tau^{\rm num}_{\rm esc}$ is obtained by fitting an
exponential to the tail of $P_{1} (\tau_{p})$. In this paper we
attempt to obtain $\tau_{\rm min}$ and $\tau_{\rm esc}$ theoretically,
as opposed to the numerical estimates $\tau^{\rm num}_{\rm min}$ and
$\tau^{\rm num}_{\rm esc}$ just described.  Note that the quantities
$\tau^{\rm num}_{\rm min}$ and $\tau^{\rm num}_{\rm esc}$ are
analogous to the quantities $\tau_e$ and $\tau_a$ respectively as
discussed in \cite{Pikovsky}.

%An almost similar distribution has been given by Lacasta et. al. 
%\cite{Lacasta} for explaining their phenomenon of anticoherence resonance 
%and in place of $\tau_{min}$ which we use, they used $t_{1}$ which was 
%associated with the characteristic time of the first maximum of the 
%relaxational oscillations.

%The Probability of $\tau_{p}$ is given by: 
%\begin{equation}
%\mathbb{P}(\tau_{p}) = \frac{1}{\tau_{esc}(D)}\Theta{(\tau_{p} -\tau_{min}(D))}
%\exp(-\frac{(\tau_{p} - \tau_{min}(D))}{\tau_{esc}(D)}) 
%\label{eqprobdist}
%\end{equation}
 
The first and the second moments of $\tau_p$, namely $\langle \tau_{p}\rangle$
and $\langle\tau_{p}^{2}\rangle$, can be easily obtained using Eq.~(\ref{distri_taup}) 
%\begin{eqnarray}
%<\tau_{p}> %= \int_{0}^{\infty}\tau_{p}\mathbb{P}(\tau_{p}) d\tau_{p} 
%&=& \tau_{min}(D) + \tau_{esc}(D)
%\label{mean} \\
%<\tau_{p}^{2}> &=& \tau_{min}^2(D) + 2\tau_{min}(D)\tau_{esc}(D) + 
%                 2\tau_{esc}^2(D).
%\label{2ndmom}
%\end{eqnarray}
and using them in the definition of $V_{N}$ we get 
%turn, from Eq. (\ref{snr}) we get 
\begin{equation}
V_{N} = \frac{\tau_{\rm esc}(D)}{\tau_{\rm min}(D) + \tau_{\rm esc}(D)}.
\label{NV}
\end{equation}
The simple formula for $V_{N}$ above, is the central result of this paper
\cite{foot1}, and is a good approximation in general for any non-linear system
exhibiting coherence resonance, provided one could predict $\tau_{\rm
esc}(D)$ and $\tau_{\rm min}(D)$ theoretically.  In what follows we
try to do the latter. A similar formula as Eq. (\ref{NV}) 
was derived, although in the low $D$ limit \cite{Lacasta} for 
anti-coherence resonance.

Formally, the resonance point is obtained by setting the derivative
of $V_{N}$ w.r.t. $D$ equals $0$. That implies the following relation
\begin{equation}
\tau^{\prime}_{\rm esc}(D_{res})\tau^{}_{\rm min}(D_{res})  = \tau^{}_{\rm esc}(D_{res}) \tau^{\prime}_{\rm min}(D_{res}). 
\label{Dres}
\end{equation}
Here $D_{res}$ denotes the value of D at the minimum of the $V_{N}$ curve i.e. at 
the resonance point. $\tau^{\prime}_{\rm esc}(D_{res})$ and $\tau^{\prime}_{\rm min}(D_{res}) $ denote their respective derivatives with $D$ evaluated at
$D_{res}$ . However, since both $\tau_{\rm min}(D_{res})$ and $\tau_{\rm esc}(D_{res})$ are system specific and are obtained partly numerically, the scope of 
the analytical application of Eq.~(\ref{Dres}) is limited.

We start with a hypothesis about the functional dependence of
$\tau_{\rm min}$ on $D$.  We claim that the action of noise on
Eq. (\ref{fhn2}) (or \ref{eqn2}) merely shifts $a$ (or $v$) to $a_{\rm
eff}$ (or $v_{\rm eff}$), with $a_{\rm eff}=a_0-D$ (or $v_{\rm
eff}=v_0-D$). To be brief let us focus on the FHN system and the
parameter $a$. The parameter value $a_0$ corresponds to the initial
stable fixed point. The $a_{\rm eff}$ makes the system feel that it is
on the LC side, across the bifurcation threshold $a_{\rm th}=1$, and
lead to a spike. The width of the spike $\tau_{\rm min}$ is expected
to be equal to the time period of the effective LC experienced, say
$t_{\rm lcp}$, i.e.,
\begin{equation}
\tau_{\rm min}(D)=t_{\rm lcp}(a_{\rm eff}),\;\mbox{and analogously for}\;v.
\label{tau_min}
\end{equation}
Here we assume that $t_{\rm lcp}$, which is the property of the system
is known apriori as a function of $a$. Note that the system can spike
even if $a_{\rm eff}$ does not cross $a_{\rm th}$ (and $\tau^{\rm
num}_{\rm min}$ can be measured numerically), but our above claim is
not valid as $t_{\rm lcp}$ is undefined.  In the later case, we
would claim that $\tau_{\rm min}(D) = w_{\rm s}$, 
the spike width.
%Surprisingly a bold claim
%can be made about it from the following simple argument. For the
%values of the parameter (say) $b$ for which a system has a limit
%cycle, the time period of each cycle $t_{\rm {lcp}}(b)$ is a specific
%function of $b$, characteristic of that system. Now our claim is that
%an addition of random noise $D$ to the parameter $b$ makes the system
%{\it feel an effective parameter value of} $b_{\rm eff} = b_0 - D$ 
%{\it which lies in the limit cycle region}.  
%Here $b_0$ is the steady state set-point of $b$ above the bifurcation
%threshold $b_{\rm th}$. For example, for FHN, $b=a$ and $a_0$ is
%slightly above $a_{\rm th} = 1$, while for CO, $b=v$ and $v_0$ is
%slightly above $v_{th} = 29.235$. Notationally, $b_{\rm eff} = a_{\rm eff}$
%and $v_{\rm eff}$, respectively, for FHN and CO. It immediately follows that  
%\begin{equation}
%\tau_{\rm min}(D) = t_{\rm {lcp}}(b_{\rm eff}) = t_{\rm {lcp}}(b_0 - D).
%\label{tau_min}
%\end{equation}
%Here we are assuming the $t_{\rm lcp}$ for any system is known {\it a priori}. 
%Further, the above formula makes sense only if $t_{\rm {lcp}}$ can be
%defined for its argument, and that is true only if $b_{\rm eff} \leq
%b_{\rm th}$, i.e. $D \geq (b_0 - b_{\rm th})$. Thus for the moment we
%have no theory for $\tau_{\rm min}(D)$ for $D < (b_0 - b_{\rm th})$,
%5although $\tau^{\rm num}_{\rm min}$ can be measured. 

\begin{figure}
\includegraphics[scale=0.33]{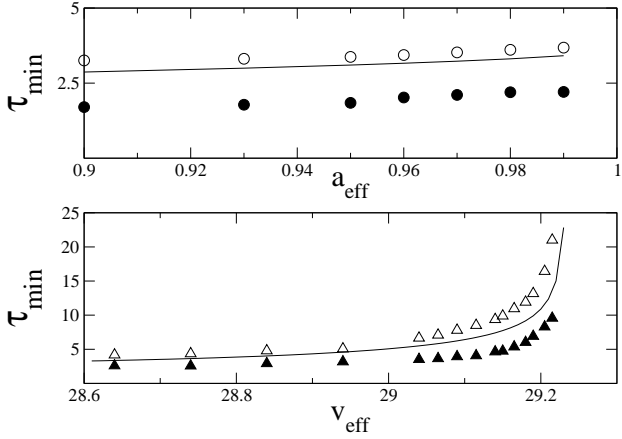}
\caption{Top frame: $\tau_{\rm min}(D)$ versus $a_{\rm eff}(D)$ in
FHN.  Bottom frame: $\tau_{\rm min}(D)$ versus $v_{\rm eff}(D)$ in
CO. Here $v_{0}= 29.24$, $\epsilon = 0.03$, $\alpha = 0.1$,
$a_{1}=1.125$, $a_{2}=-0.075$, $a_{3} = 0.00125$, $R = 10$ (see
Eq. (\ref{eqn2})) --- these same values are used for CO in other
figures. For both frames: The empty symbols are for $\tau^{\rm
num}_{\rm {min,r}}$ and filled symbols are for $\tau^{\rm num}_{\rm
{min,l}}$.  The solid lines represent $\tau_{\rm min}$ from
Eq.~(\ref{tau_min}).  The numerical data and the theoretical curves
shows excellent agreement.}
\label{tau_min_fig}
\end{figure}

We proceed to test Eq.~(\ref{tau_min}) in FHN and CO models. In both
the top (for FHN) and bottom (for CO) frames of
Fig. \ref{tau_min_fig}, the solid lines are as per
Eq.~(\ref{tau_min}). Instead of plotting $\tau^{\rm num}_{\rm min}$,
for more clarity, we have plotted $\tau^{\rm num}_{\rm {min,r}}$ in
empty symbols and $\tau^{\rm num}_{\rm {min,l}}$ in filled symbols. 
The fact that $\tau_{\rm min}$ falls in between $\tau^{\rm
num}_{\rm {min,r}}$ and $\tau^{\rm num}_{\rm {min,l}}$ for the range
of $D$ studied, and the agreement being excellent for two distinct systems
FHN and CO (with distinct $t_{\rm {lcp}}(a)$ and $t_{\rm {lcp}}(v)$
functions), gives strong empirical support for the formula in
Eq.~(\ref{tau_min}).

Next, we turn to $\tau_{\rm esc}$ in Eq.~(\ref{NV}). The dynamics of
one of the variables in the non-linear system, for example $y$ in FHN
or $c$ in CO, under finite noise strength $D$, can be viewed as a
stochastic process around the stable fixed points $y_*$ or $c_*$,
respectively. For subsequent discussion we focus on FHN, but the
results apply generally to any non-linear system exhibiting coherence
resonance.  Most often the noise displaces $y$ a little and then it
relaxes back to the fixed point, in a typical excursion time
$\bar{\tau_0}$. Occasionally however, if the excursion of the variable
(e.g. ${\Delta y} = y-y_*$ in FHN) falls below a certain threshold
denoted by a typical $-\delta_m$ (here $\delta_m > 0$), the system
exhibits a cycle and $y$ exhibits a spike. The latter amounts to
absorption of ${\Delta y}$ at the boundary $-\delta_m$, on its first passage. 

Specific system dependent details of the shape of the effective
trapping potential is necessary to analytically calculate the above
mentioned typical first passage time $\tau_{\rm esc}$. Since our
purpose is to remain as general as possible, we make a simplifying
general assumption that {\it after every random kick the relaxation is
instantaneous}. In effect this is equivalent to coarse-graining in
time over units of the typical excursion time $\bar{\tau_0}$ (mentioned
above and defined below). 
%For $\tau_{esc}(D)$ :- 
\begin{figure}
\includegraphics[scale=0.32]{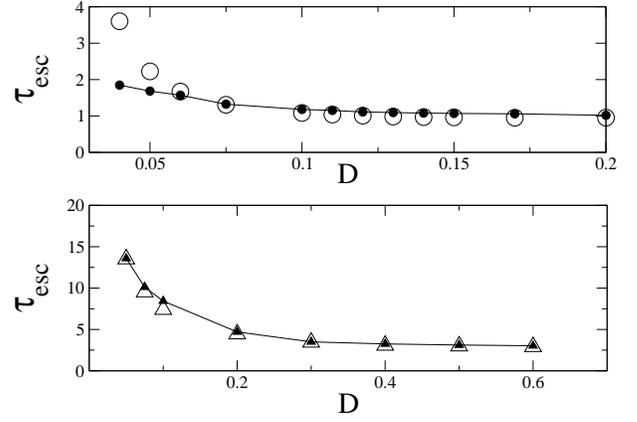}
\caption{$\tau_{\rm esc}(D)$ versus $D$ for FHN (top frame) and CO (bottom 
frame) denoted by line joining filled symbols (the values of $\bar\tau_{0}(D)$
and $\delta_m(D)$ used for this plot are discussed in the text and 
fig.~\ref{taudelta}. Open symbols represents the numerical values 
$\tau_{\rm esc}^{\rm num}(D)$, obtained from fits as in fig.\ref{probdist}.} 
\label{tesc}
\end{figure}

%It implies 
%that the stochastic 
%process is just the opposite of free random walk, and after every
%noise kick, ${\Delta y}$ returns to $0$ instantaneously. 
Thus every excursion ${\Delta y} \neq 0$ at every discrete time step,
may be treated as independent, and merely follows the noise and
therefore has the same (Gaussian) distribution as the noise. Then it
immediately follows, that the probability $Q(n)$ that the signal
${\Delta y}$ does not go below $-\delta_m$ for $n$ successive time
steps and does so in the $(n+1)^{\rm th}$ step is
\begin{eqnarray}
Q(n) &=& [{P}_{>}(-\delta_m)]^{n} {{P}_{<}(-\delta_m)} \nonumber \\
 &=& e^{-n \ln{(\frac{1}{P_{>}(-\delta_m)})}}{{P}_{<}(-\delta_m)}
\label{1st_pass_prob} 
\end{eqnarray}
where  ${P}_{>}(y) = \frac{1}{\sqrt{\pi D^{2}}} 
{\int_{y}^{\infty}}e^{-\frac{x'^{2}}{D^{2}}} dx'$ and $P_{<} = 1 - P_{>}$. 
Eq~(\ref{1st_pass_prob}) shows that $Q(n)$ is exponential distributed,
and its decay constant gives the ``typical first passage time'' 
\cite{gardiner} in units of $\bar{\tau_{0}}$:    
\begin{eqnarray}
\tau_{\rm esc}(D)/\bar\tau_{0} &=& -[{\ln{({P}_{>}(-\delta_m))}}]^{-1},
~~~{\rm where}~ \nonumber \\ 
{P}_{>}(-\delta_m) &=& (1 + {\rm erf}({\delta_m}/D))/2  
\label{1st_pass_time}
\end{eqnarray}
and ${\rm erf}(.)$ is the Error function \cite{arfken}.

%For $\tau_{0}(D)$ :-
\begin{figure*}[t]
\includegraphics[scale=0.7]{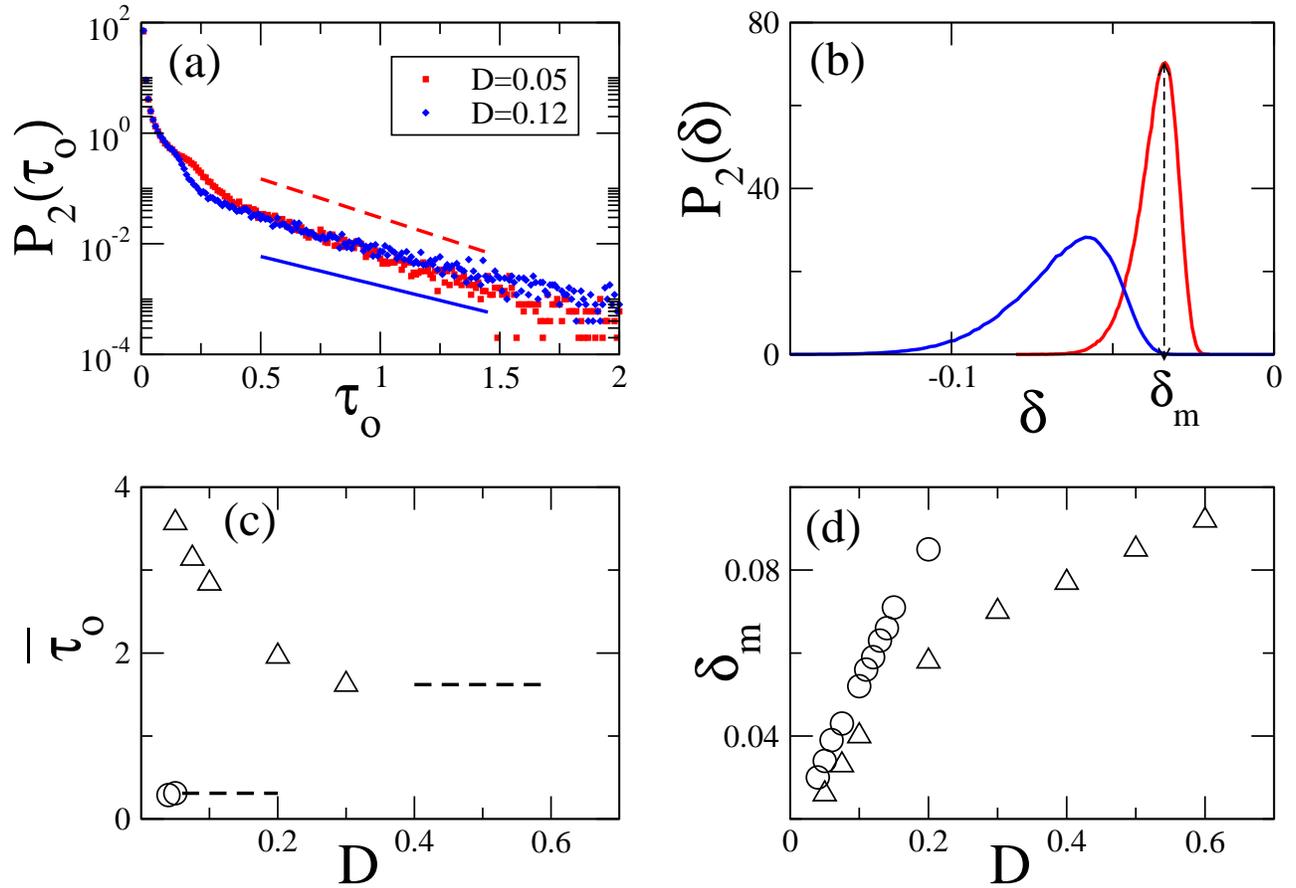}
\vspace{.2cm}
\caption{(Color online) (a) PD of $\tau_{0}$ for the FHN along with
exponential fit $e^{-t/\bar{\tau_{0}}}$ for two different $D$
values. (b) PD of $\delta$, and depiction of $\delta_m$ as most
probable $\delta$ for the same two $D$ as in (a). (c) $\bar{\tau_{0}}$
versus $D$ for FHN ($\circ$) and CO ($\bigtriangleup$) extracted from
figure (a). The dashed line segments are explained in the text. (d)
$\delta_{m}$ versus $D$ extracted from figure (b) -- two sets are for
FHN ($\circ$) and CO ($\bigtriangleup$).}
\label{taudelta}
\end{figure*}
 
Note that the $D$ dependence of $\tau_{\rm esc}$ comes from explicit
dependence of $P_{>}$ on $D$, as well as the implicit dependence of
the time unit $\bar\tau_{0}$ and barrier location $\delta_m$ on $D$.
Of course $\bar\tau_{0}$ and $\delta_m$ will be system specific and
incorporate the detail nature of the dynamic potential trap.
The procedure to find $\bar\tau_{0}(D)$ and $\delta_m(D)$ will be 
discussed later.  If we assume that the latter two quantities are
known {\it a priori} then Eq.~(\ref{1st_pass_time}) maybe claimed to be
a ``theoretical'' formula, and compared to the numerical values of
$\tau^{\rm num}_{\rm esc}$ obtained as in Fig.~\ref{probdist}. In
Fig.~\ref{tesc} we see that the agreement between the theoretical
formula and numerical data are excellent.

%From the definition just below Eq.~(\ref{1st_pass_prob}), few steps of
%algebra leads to $P_{>}(-\delta_m) = (1 + {\rm erf}({\delta_m}/D))/2$
%(where ${\rm erf}(.)$ is the Error function). 
Using the asymptotic expansion of ${\rm erf}(.)$ \cite{arfken} in
Eq.~(\ref{1st_pass_time}) we get $\tau_{\rm esc}/\bar\tau_{0} \approx
{\rm constant}$ for $D/{\delta_m} \gg 1$ and $\approx
e^{{\delta_m^2}/{D^2}}$ for $D/{\delta_m} \ll 1$. The latter behavior
has been referred to as Kramer's formula for $\tau_{\rm esc}$
\cite{Pikovsky, Lacasta}, but one needs to be careful --- unlike the
usual Kramer's escape time formula, $\delta_m$ is not the barrier
height of the potential well but rather proportional to the width of
the well.

What remains to be discussed is determination of $\bar{\tau_{0}}(D)$
and $\delta_m(D)$. To define $\bar{\tau_{0}}$ precisely, we note that 
between two successive spikes of $y$, 
%(see Fig.~\ref{timeser}(a))
the process $\Delta y$ (and $\Delta c$ for CO model) crosses zero several 
times. Let $\tau_0$ be the time interval between zero crossings 
of $\Delta y$ which is same as the excursion time mentioned earlier. 
%then $\tau_0$ is literally the typical time for returning 
%to zero after getting kicked away from zero by random force. 
%Thus coarse-graining 
%over time interval $\tau_0$ is consistent with our initial assumption
%of instantaneous return to zero. 
A probability distribution (PD) of 
$\tau_0$ is then found for every $D$, and the PD has an exponential tail 
as shown in Fig.~\ref{taudelta}(a). We define the time constant of the 
latter exponential fit to be $\bar{\tau_{0}}(D)$. 
For FHN and CO systems 
the $\bar{\tau_{0}}(D)$ thus obtained are shown in Fig.~\ref{taudelta}(c). 
But with increasing $D$ the time stretches between two spikes become very small,
%(see Fig.~\ref{timeser}(b))
making determination of $\bar{\tau_{0}}(D)$ unreliable due to poor statistics. 
So we took $\bar{\tau_{0}}(D)$ to be a constant (denoted by dashed line 
segments in Fig. ~\ref{taudelta}(c)), for the $D$ values beyond which
$\bar{\tau_{0}}(D)$ could not be reliably determined.  A {\it
posteriori} justification of the latter adhoc assumption for
$\bar{\tau_{0}}$ lies in the successful agreement with numerical data
of $\tau_{\rm esc}(D)$ (see Fig.~\ref{tesc}). 
%Whatever is done for
%$\Delta y$ in FHN is similarly done for $\Delta u$ in CO.

Next, we define $-\delta$ as the threshold of $\Delta y$ at which
spiking occurs. Then the PD of $\delta$ for every $D$ can be computed
(see Fig.~\ref{taudelta}(b)) and the most probable value may be
identified as $\delta_m$. Plot of $\delta_m$ is shown
against $D$ for both FHN and CO systems in
Fig.~\ref{taudelta}(d). These values of $\delta_m$ were used to obtain the
theoretical curve in Fig.~\ref{tesc}. 

%\begin{figure}
%\includegraphics[scale=0.35 ]{fig5.eps}
%\caption{Time series of pulses in FHN for (a) D=0.05 and (b) D=0.12}
%\label{timeser}
%\end{figure}

%\begin{figure}
%\includegraphics[scale=0.35]{fig6.eps}
%\vspace{.1cm}
%\caption{Normalized Variance Plot for FHN(top) and CO(bottom). Open symbol
%denote the $V_{N}$ from numerical time series analysis. While, filled symbols 
%joined by line represents the theoretical values from Eq.~\ref{NV}.}
%\label{NVpic}
%\end{figure}
 
\begin{figure} 
\includegraphics[scale=0.35]{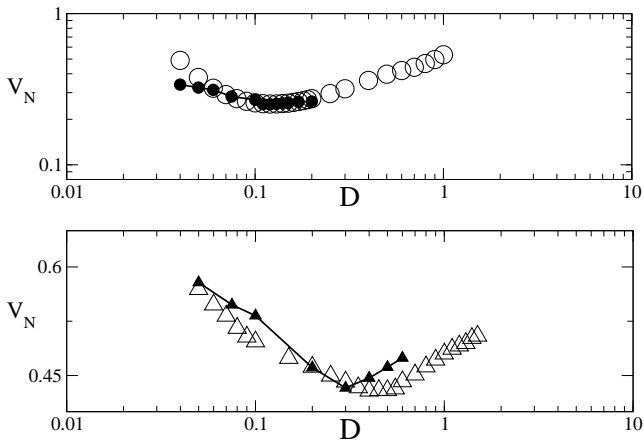}
\vspace{.1cm}
\caption{Normalized Variance Plot for FHN(top) and CO(bottom). Open symbol
denote the $V_{N}$ from numerical time series analysis. While, filled symbols 
joined by line represents the theoretical values from Eq.~(\ref{NV}). The
semi-analytic treatment was found to be valid for the following range 
of $D$ values: For the FHN model $0.04 \le D \le 0.2$. For the CO model 
$0.05 \le D \le 0.6$.}
\label{NVpic}
\end{figure}

Finally, one can directly plot the $V_{N}$ from the theoretical formulas in
Eqs. (\ref{NV}), (\ref{tau_min}), and (\ref{1st_pass_time}) and compare it
with numerical $V_{N}$ obtained from time series analysis (see Fig.~\ref{NVpic}).  
Both for FHN and CO the agreement is quite good and
the locations of resonance (the minima) are obtained within acceptable
error limits.  

\section{Conclusion}
Thus we claim to have found an alternate way of determining $V_{N}$ for
non-linear systems exhibiting coherence resonance, based on
theoretical considerations rather than brute force time series
analysis. Only three empirical inputs are required for a specific
system, namely (i) the limit cycle period $t_{\rm {lcp}}(a)$ as a
function of the control parameter $a$, (ii) the typical zero-crossing
time interval $\bar{\tau_{0}}(D)$ of the relevant stochastic dynamical
variable, and (iii) the typical distance of excursion $\delta_m(D)$
beyond which the variable maybe regarded as ``absorbed'' (i.e.  it
spikes). We highlight the fact that the effective barrier parameters
$\bar{\tau_{0}}$ and $\delta_m$ turn out to be $D$ dependent.  It may
seem no less work to obtain the empirical inputs (i)--(iii) for a
system, yet once obtained they can be substituted in the simple
theoretical formulas Eqs.~(\ref{NV}), (\ref{tau_min}), and
(\ref{1st_pass_time}) and coherence resonance maybe predicted.

\hspace{0.2cm}

%\begin{acknowledgments}
%\end{acknowledgments}

%\bibliographystyle{h-physrev3}
%\bibliography{SR4santi}

\begin{thebibliography}{10}
\bibitem{foot1}
Note that in Eq.~(\ref{NV}) if $\tau_{min} = 0$, then $V_{N} = 1$ .
\end{thebibliography}

\end{document}